\documentclass[superscriptaddress,showpacs,showkeys,amsmath,amssymb,
aps,prb,reprint,
]{revtex4-2}

\usepackage{graphicx}

\usepackage{dcolumn}
\usepackage{bm}
\usepackage{hyperref}
\usepackage[svgnames]{xcolor}
\usepackage{soul}
\usepackage[mathlines]{lineno}
\usepackage{miller}
\usepackage[colorinlistoftodos]{todonotes}

\usepackage{siunitx}
\DeclareSIUnit\oersted{Oe}
\DeclareSIUnit\emucc{\text{emu}\per\centi\meter^3}

\newcommand{\chromia}{C\lowercase{r}$_2$O$_3$}

\begin{document}

\title{Quadratic magnetoelectric effect during field cooling in sputter grown Cr$_2$O$_3$ films 
}
\author{Muftah Al-Mahdawi}
\email{mahdawi@tohoku.ac.jp}
\affiliation{Center for Science and Innovation in Spintronics (Core Research Cluster), Tohoku University, Sendai, 980-8577, Japan}
\affiliation{Center for Spintronics Research Network, Tohoku University, Sendai 980-8577, Japan}

\author{Tomohiro Nozaki}
\email{nozaki.tomohiro@aist.go.jp}

\affiliation{Research Center for Emerging Computing Technologies, National Institute of Advanced Industrial Science and Technology (AIST), Tsukuba 305-8568, Japan}

\author{Mikihiko Oogane}
\affiliation{Center for Science and Innovation in Spintronics (Core Research Cluster), Tohoku University, Sendai, 980-8577, Japan}
\affiliation{Center for Spintronics Research Network, Tohoku University, Sendai 980-8577, Japan}\affiliation{Department of Applied Physics, Tohoku University, Sendai 980-8579, Japan}

\author{Hiroshi Imamura}
\affiliation{Research Center for Emerging Computing Technologies, National Institute of Advanced Industrial Science and Technology (AIST), Tsukuba 305-8568, Japan}

\author{Yasuo Ando}
\affiliation{Center for Science and Innovation in Spintronics (Core Research Cluster), Tohoku University, Sendai, 980-8577, Japan}
\affiliation{Center for Spintronics Research Network, Tohoku University, Sendai 980-8577, Japan}\affiliation{Department of Applied Physics, Tohoku University, Sendai 980-8579, Japan}

\author{Masashi Sahashi}
\affiliation{School of Engineering, Tohoku University, Sendai 980-8579, Japan}

\begin{abstract}
    \chromia{} is the archetypal magnetoelectric (ME) material, which has a linear coupling between electric and magnetic polarizations. Quadratic ME effects are forbidden for the magnetic point group of \chromia{}, due to space-time inversion symmetry. In \chromia{} films grown by sputtering, we find a signature of a quadratic ME effect that is not found in bulk single crystals.  We use Raman spectroscopy and magetization measurements to deduce the removal of space-time symmetry, and corroborate the emergence of the quadratic ME effect. We propose that meta-stable site-selective trace dopants remove the space, time, and space-time inversion symmetries from the original magnetic point group of bulk \chromia{}. We include the quadratic ME effect in a model describing the switching process during ME field cooling, and estimate the effective quadratic susceptibility value. The quadratic magnetoelectric effect in a uniaxial antiferromagnet is promising for multifunctional antiferromagnetic and magnetoelectric devices that can incorporate optical, strain-induced, and multiferroic effects.
\end{abstract}
\date{\today}

\maketitle
\section{Introduction}
The coupling between the magnetic and electric orders in single-phase materials is interesting for the fundamental physics research, and applications in multi-functional multi-input information storage and processing devices \cite{pyatakov_2012}. The magnetoelectric (ME) effect arises in the expansion of the thermodynamic potential as coupling terms between the electric $E$ and magnetic $H$ fields \cite{dzyaloshinskii_1960}. The linear ME effect was theoretically predicted and experimentally observed first in the antiferromagnet corundum-type \chromia{} \cite{dzyaloshinskii_1960,astrov_1960,astrov_1961}, and later found a renewed interest in ME-type memories \cite{borisov_2005,ashida_2014,toyoki_2015a}. Higher-order ME effects that are quadratic in $E$ or $H$ have been found in various materials \cite{cardwell_1971,takano_1991,fogh_2020, liang_2011, weymann_2020}.
Equivalent to quadratic ME effects, the linear magneto-optical effect was also found for other antiferromagnets \cite{kharchenko_1994}. However, the simultaneous presence of quadratic and linear ME effects in the same material phase is not frequently found.  A necessary condition of the linear ME effect is the symmetry breaking of a magnetic crystal under space-inversion ($I$) and time-reversal ($R$) operations \cite{odell_1970a,borovik-romanov_2013}. The quadratic ME effects require a breaking of space-time inversion ($I\!R$) symmetry and either of $I$ or $R$ symmetries \cite{schmid_1973, borovik-romanov_2013}. The understanding of interplay between linear and nonlinear ME effects will be of interest for applications in ME memories and devices.

\begin{figure}
	\includegraphics[width=0.46\textwidth]{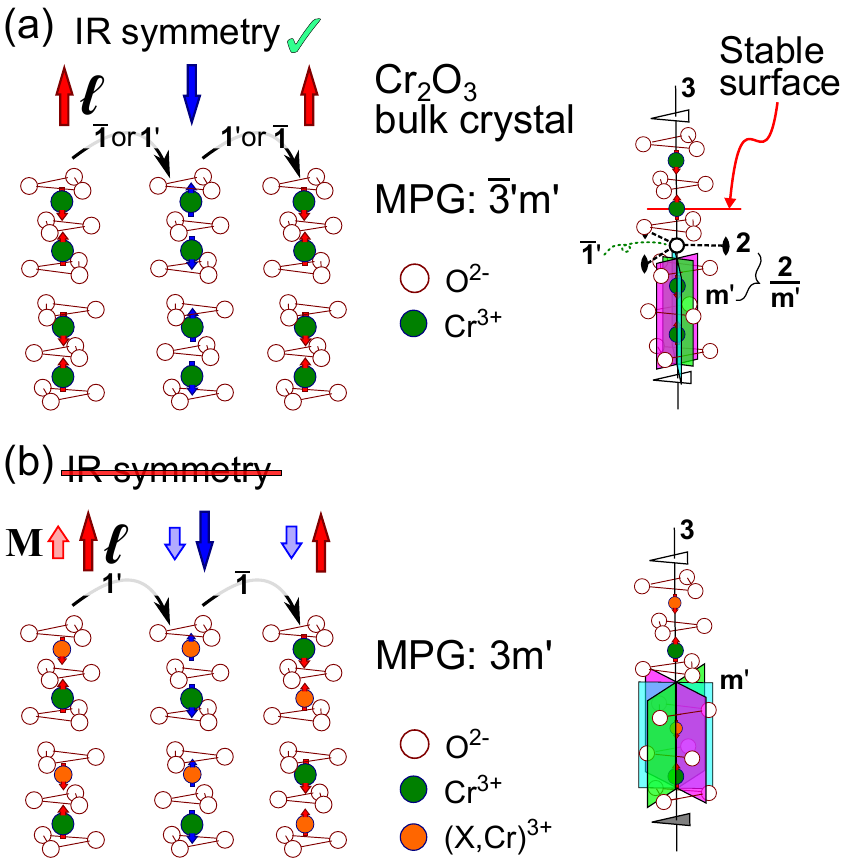}
	\caption{(a) The spin configuration of Cr$_2$O$_3$ breaks either the $I\!\equiv\!\Bar{1}$ or $R\!\equiv\!1'$ symmetries, but it is symmetric under the combined $I\!R\equiv\!\Bar{1}'$ operation. (b) We propose that the presence of non-equivalent spin moments breaks the combined $I\!R$ symmetry, which results in an uncompensated magnetization $M$ and quadratic ME effects. Schematics of magnetic point groups $\bar{3}'m'$ and $3m'$ for each case are shown on the right side.
	\label{fig:mpg}
	}
\end{figure}

  \chromia{} has the same rhombohedral crystal structure as corundum $\alpha$-Al$_2$O$_3$, with the crystallographic point group of $\bar{3}m$. Cr$^{3+}$ cations fill $2/3$ of distorted O$^{2-}$ anions octahedra. The Cr cations lie along the $\bar{3}$ axis, with an inversion center in an empty octahedron, and the O anions lie on the $2\!/\!m$ axes.
 Above the N\'eel temperature of \chromia{} ($T_N = \SI{307}{\kelvin}$), the spin configuration is paramagnetic, and the magnetic point group (MPG) is the gray $\bar{3}m1'$, where the individual $I$ and $R$ symmetries are present, similar to the parent corundum rhombohedral crystal.
Upon the magnetic ordering below $T_N$, the MPG is lowered to $\bar{3}'m'$, which breaks the single $I$ and $R$ symmetries \cite{dzyaloshinskii_1960}. There are two distinct spin configurations of $\downarrow \uparrow\! \cdot\! \downarrow \uparrow$ ($L^+$ domain state) and $\uparrow \downarrow\!\cdot\!\uparrow \downarrow$ ($L^-$ domain state), where the dot denotes the inversion center. The order parameter is the antiferromagnetic order vector, defined from the four Cr spins as $\ell = s_1 - s_2 + s_3 - s_4$ [Fig.~\ref{fig:mpg}(a)]. Both of $\ell$ and the linear ME susceptibility $\alpha$ change signs under $I$ and $R$ operations. In the convention used by Birss \cite{birss_1964,grimmer_1991}, both of $\ell$ and $\alpha$ are $-c$ tensors, and they are commensurate with each other.

The ME effect can be expanded with higher order terms. The quadratic effects are represented by the energy terms $\beta E H^2$ for electrobimagnetic effect, and $\gamma H E^2$ for magnetobielectric effect. The third-rank susceptibility tensors $\beta$ and $\gamma$ have nonvanishing components in MPGs that permit piezoelectricity and piezomagnetism, respectively \cite{ascher_1968}.
The MPG $\bar{3}'m'$ of \chromia{} has the combined $I\!R$ symmetry \cite{dzyaloshinskii_1960}, as shown in Fig.~\ref{fig:mpg}(a). The presence of $I\!R$ symmetry permits the linear ME effect, but forbids the quadratic ME effects, and the various pyro-/ferro-/piezo- electric and magnetic effects.This condition can be seen in the effect of $I\!R$ operation on the relevant ME energy terms in the expanded thermodynamic potential energy $F$, as follows:

\begin{align}
    &F = -\alpha E H -\frac{1}{2} \beta E H^2 - \frac{1}{2} \gamma E^2 H + \cdots , \nonumber \\
    &(R\circ I)(F) - F = + \beta E H^2 +  \gamma E^2 H \, .
    \label{eq:F}
\end{align}

The presence of $I\!R$ symmetry imposes $(R\circ I)(F) - F = 0$, and the quadratic ME effects are prohibited. For more details and forms of tensors, we refer to the reviews in Refs.~\cite{schmid_1994,schmid_2008,borovik-romanov_2013,kopsky_2015}, and the original works in Refs.~\cite{ascher_1968,schmid_1973,mercier_1977,grimmer_1991,grimmer_1994}. 

In this work, we found an evidence of a quadratic ME effect in sputtered \chromia{} films, which is not permitted in the MPG $\bar{3}'m'$ of bulk \chromia{}. We propose that non-equivalent spin moments at each sublattice removes the $I\!R$ symmetry, and the MPG will be of $3m'$ [Fig.~\ref{fig:mpg}(b)]. First, we clarify the lack of $I$, $R$, and $I\!R$ symmetries from measurements of the thermoremnant magnetization and Raman spectroscopy. 
After that, we investigate the quadratic ME effect from measurements on the domain switching probability during the ME field cooling (MEFC) process. Then, we use this procedure to estimate the relative magnitude of the quadratic ME effect in our sample.

\section{Experimenal details}
We used samples similar to earlier reports \cite{al-mahdawi_2017a, al-mahdawi_2017b, nozaki_2017a, nozaki_2018b}.
The reference sample is a $c$-cut bulk single-crystal substrate, having dimensions of $4\times 4 \times 0.5$ \SI{}{\milli\meter^3}, prepared by the Verneuil process and acquired commercially. 
The film samples had the stack structure of Al$_2$O$_3$-\hkl(0001) substrate/Pt \SI{25}{\nano\meter}/\chromia{}-\hkl(0001) \SIlist{500;1000;2300}{\nano\meter}/no capping, where the numbers are the layer nominal thicknesses. The Pt layer was grown by \emph{dc} magnetron sputtering, and the \chromia{} film layer was grown by reactive \emph{rf} magnetron sputtering from a metallic Cr target (\SI{99.99}{\percent} purity) in a mixture of argon and oxygen gases. The growth substrate temperature of \chromia{} and Pt layers was kept relatively low at \SI{773}{\kelvin}, to preserve the meta-stable structure. 

The determination of the symmetry elements by crystal diffraction techniques is unfeasible. Xray or neutron-based diffraction methods determine only the Laue class, and cannot determine the presence or absence of $I$ symmetry in our \chromia{} films (Sec.~S1 in \cite{NoteSupp}). 
On the other hand, vibrational spectra are sensitive to $I$ symmetry of the crystallographic point group. Also, magnetization and magnetoelectric effects are determined by the $R$ and $I\!R$ symmetries. We measured the thermoremnant magnetizations and Raman spectra to infer which symmetries are removed from the original \chromia{} MPG of $\Bar{3}'m'$.

We characterized the thermoremnant magnetic properties by a SQUID magnetometer, where we measured the magnetization along the out-of-plane direction, parallel to the $c$-axis. After a field cooling in a magnetic field $H_\mathrm{fr}=\SI{10}{\kilo\oersted}$ from \SI{330}{\kelvin} down to \SI{10}{\kelvin}, we set the magnetic field to zero, and we measured the magnetization in the heating direction at \SI[per-mode = symbol]{1}{\kelvin\per\minute} heating rate.
We measured Raman scattering in the same samples at room temperature $\approx \SI{293}{\kelvin}$. We used a micro-Raman spectrometer, equipped with a green laser ($\lambda = \SI{532.133}{\nano\meter}$, \SI{50}{\milli\watt}), and measured in the $\Bar{Z}(YY)Z$ back scattering geometry configuration. We set the span of Raman shift at \SIrange{200}{2500}{\per\centi\meter}, with a resolution of \SI{6.8}{\per\centi\meter}. The Pt bottom layer blocks the laser beam, and no Raman scattering was detected from the sapphire substrate.

We measured the ME properties of two other samples. The first is a bulk single crystal metallized on both \hkl(0001) faces by Ta \SI{10}{\nano\meter}/Cu \SI{300}{\nano\meter} electrodes. The second is Al$_2$O$_3$ substrate/Pt \SI{25}{\nano\meter}/\chromia{} \SI{500}{\nano\meter}/Pt \SI{25}{\nano\meter} fabricated into a $4\times 2 $ \SI{}{\milli\meter^2} cross capacitor structure. 
We characterized the average domain configurations by the converse ME effect, where the electrically-induced magnetization was measured by lock-in detection in a SQUID magnetometer. More details of the measurement setup are described elsewhere \cite{al-mahdawi_2017a}.

\section{Symmetry in sputtered \chromia{} films}

\begin{figure}
   	\includegraphics[width=0.46\textwidth]{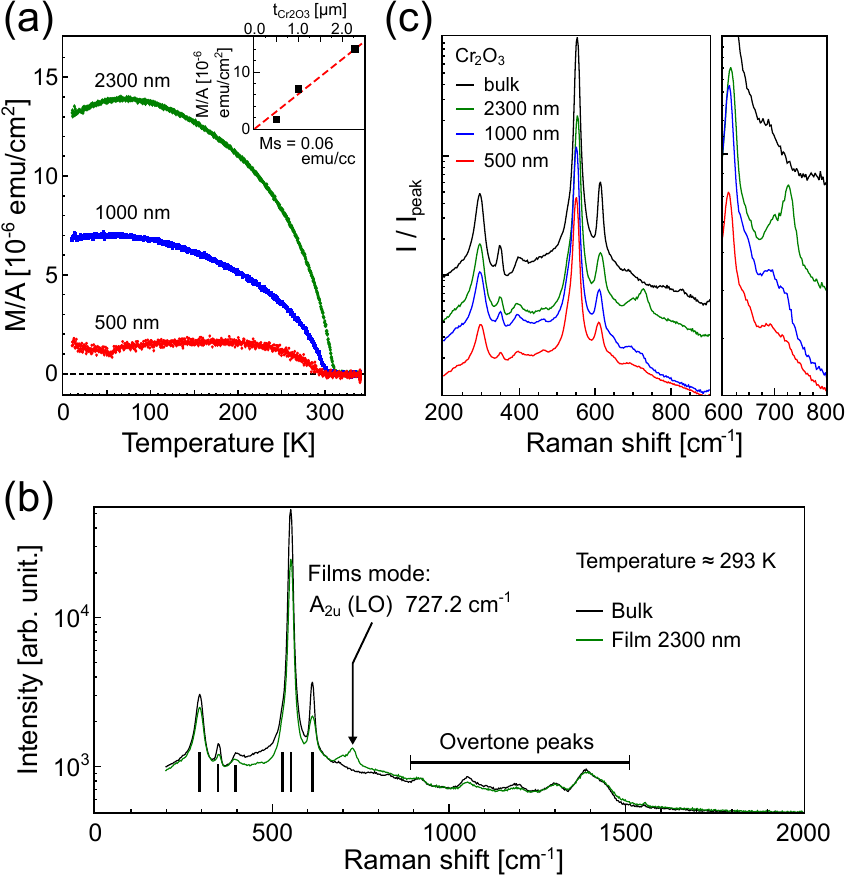}
	\caption{(a) Temperature-dependence of thermoremnant magnetiztion in the sputtered \chromia{} films. The inset shows the scaling of areal magnetization density with thickness. (b) The wide-span Raman spectra of the bulk and 2300-nm samples. The location of Raman modes are listed in table \ref{tab:modes}. (c) Comparison of the Raman spectra. An enlarge view on the \SI{727.2}{\per\centi\meter} mode is on the right panel.
	} 
	\label{fig:raman_M}
\end{figure}

Fig.~\ref{fig:raman_M}(a) shows the thermoremnant magnetization measurements, at zero field after field cooling. The temperature dependence shows pyromagnetism, and the transition temperature coincides with the $T_N$ of \chromia{}. The magnetization areal density scales linearly with the film thickness [inset of Fig.~\ref{fig:raman_M}(a)], showing that the magnetization is within the volume of the films. The estimated low-temperature magnetization value is very low at \SI{0.06}{\emucc}.

A finite magnetization in \chromia{} films was reported before. It was attributed to uncompensated moments at the films surface, either by interfacial misfit dislocations \cite{kosub_2017}, or by the boundary magnetization due to removal of $I$ symmetry at the surface \cite{he_2010,fallarino_2014}. In contrast, we found that sputtered \chromia{} films have volume pyromagnetism, likely of a ferrimagnetic-type origin \cite{al-mahdawi_2017a,al-mahdawi_2017b,nozaki_2017a,nozaki_2018b, nozaki_2020}. The magnitude of $M$ and the sign of $\ell \cdot M$ can be controlled by dopants concentration and type. For example, we found that aluminium doping results in a parallel orientation between $\ell$ and $M$. At \SI{0.07}{\percent} Al doping level, we had $M = \SI{2}{\emucc}$, and 
it reached up to \SI{60}{\emucc} at \SI{3.7}{\percent} Al doping level \cite{nozaki_2018b}. In the present results, the \chromia{} films are nominally non-doped, but the pyromagnetism origin is the same. There are trace dopants present in the sputtering deposition targets or chamber walls, which are incorporated in the films. We have a volume magnetization of $\SI{0.06}{\emucc}$, which corresponds to a doping level of \SI{35}{ppm}. The high-purity sputtering targets have a guaranteed purity at \SI{100}{ppm} level. The trace elements are likely to be Al and Si, based on the origin of target materials. Therefore, even nominally non-doped films have a ferrimegnetic-type volume pyromagnetism.

 Ferro- and ferri-magnetism are allowed only if the MPG is a subgroup of $\infty\!/\!m\!m'$ \cite{schmid_2008}. Magnetic crystals with either $R$ or $I\!R$ symmetries cannot support ferromagnetism. The MPG $\bar{3}'m'$ of bulk \chromia{} has the $I\!R$ symmetry \cite{dzyaloshinskii_1960}. To remove the $I\!R$ symmetry, non-equivalent spin moments at each sublattice are required, and the MPG will be of $3m'$ [Fig.~\ref{fig:mpg}(b)].
Cation dopants in \chromia{} substitute the host Cr sites \cite{arca_2017,nozaki_2019}.
Site-selective substitution by nonmagnetic dopants reduces one of the sublattice magnetic moments on average and/or induces ion displacement by chemical pressure. 
The corundum-type crystals have a surface that is stable if terminated in the bottom half of the buckled metal-ion layer, as indicated in Fig.~\ref{fig:mpg}(a) \cite{guo_1992,rohr_1997,mejias_1999,wang_1998}. The topmost Cr ion has a large relaxation towards the oxygen plane, and the oxidation state is close to Cr$^{2+}$, instead of Cr$^{3+}$ of the bulk \cite{rohr_1997,mejias_1999}. The surface electrostatic environment is closer to a lithium niobate (LiNbO$_3$)-type environment, and the occupation energy does not become equivalent to the dopant and Cr ions. Therefore, a metastable site-selective substitution accumulates during the layer-by-layer growth of doped \chromia{} films, under relatively-low growth temperatures. 
The resulting MPG below $T_N$ becomes $3m'$, instead of $\Bar{3}'m'$ [Fig.~\ref{fig:mpg}(b)]. This is only observed for a layered film growth mechanism, and not for films prepared by post-annealing \cite{fallarino_2015}, or bulk doped-\chromia{} powders, \emph{c.f.~}supplementary information of \cite{nozaki_2018b}. In the $3m'$ MPG, the $\alpha$ tensor has the same form as the $\bar{3}'m'$. This explains why the $\alpha$ in \chromia{} films is similar to the bulk single crystal in magnitude and temperature dependence, regardless of doping \cite{borisov_2016,al-mahdawi_2017a,nozaki_2018b, nozaki_2020}.

The proposition of site-selective substitution has the $I$ symmetry removed also from the parent crystallographic point group, as seen after the $I$ operation in Fig.~\ref{fig:mpg}(b). This makes our \chromia{} films polar, even in the paramagnetic phase, similar to lithium niobate. In vibrational spectra, when an $I$-symmetry is present in the parent crystallographic point group, there is a mutual exclusion between the Raman-active and infrared(ir)-active vibrational modes. 
 In a rhombohedral corundum crystal, the irreducible representation of the optical modes is \cite{bhagavantam_1939,mougin_2001,bilbao_raman_2003,bilbao_aroyo_2006}:

\begin{equation}
        \Gamma_\mathrm{opt} = 2 A_{1g} + 5 E_g + 2 A_{2u} + 4 E_u + 2 A_{1u} + 3 A_{2g}. 
\end{equation}
The Raman-active modes are two $A_{1g}$ and five $E_g$ modes, the ir-active modes are two $A_{2u}$ and four $E_u$ modes, and three $A_{2g}$ and two $A_{1u}$ modes are silent. 
In Fig.~\ref{fig:raman_M}(b), we show a comparison between the wide-span Raman spectra of the bulk and 2300-\si{\nano\meter} samples. We note that the measurement temperature is room temperature , and the focused laser spot heats the local temperature above $T_N$. Thus, the Raman spectra are acquired in the paramagnetic phase.
All the peaks in the bulk sample can be assigned to the reference values in literature, even though there is no agreed consensus on each of them \cite{*[{}][{and references therein}] shim_2004}. The film sample has an identical spectrum for most of the peaks. The Pt buffer layer increases the collection of scattered light, and results in a high-intensity spectrum. In Fig.~\ref{fig:raman_M}(c), we show the Raman spectra acquired from the same samples of Fig.~\ref{fig:raman_M}(a), and we list the position of our measured modes in table \ref{tab:modes}. The peak positions of the films are close to the bulk values, indicating that the films are relaxed with a negligible stress \cite{birnie_1992,mougin_2001}. 

A new mode at a Raman shift of \SI{727.2}{\per\centi\meter} appears only in the film samples [identified by an arrow in Fig.~\ref{fig:raman_M}(b)]. This mode also increases in intensity with increasing film thickness [right panel of Fig.~\ref{fig:raman_M}].
 In \chromia{}, the most intense modes are the phonon modes of $A_{1g}$ and $A_{2u}$ \cite{lucovsky_1977a, mccarty_1989}. 
The films' new mode at \SI{727.2}{\per\centi\meter} can be identified with a $A_{2u}$ longitudinal-optical (LO) mode \cite{lucovsky_1977a}, which should be an ir-active Raman-inactive mode.
The site-selective substitution in \chromia{} films removes the centrosymmetry of the parent rhombohedral point group, and the rule of mutual exclusion does not hold. The $A_{1g}$ and $A_{2u}$ modes will have $A_1$-symmetry character, and both become Raman-active. Similar findings of activation of ir-only-modes were found in Cr$_2$O$_3$-Fe$_2$O$_3$ solid solutions \cite{mccarty_1989}, and in pure \chromia{} at ambient pressure after applying \SI{61}{\giga\pascal} \cite{shim_2004}, and they were attributed to symmetry lowering.
Two additional weak peaks appear for the films at \SIlist{462;693}{\centi\meter^{-1}}. The positions of these modes coincide with $A_2$-symmetry modes from first-principles calculations \cite{larbi_2017}. When centrosymmetry is removed, the silent $A_{1u}$ and $A_{2g}$ modes become $A_2$ in symmetry, which is still silent for both of Raman and ir spectroscopy. However, the $A_2$ modes are active for the two-photon hyper-Raman scattering, which is a significantly weak scattering process. A possible explanation is that resonant enhancement of hyper-Raman scattering occurs when the two-photon pumping energy aligns with an electronic transition \cite{chung_1988,kelley_2010}.
The peaks at 916, 1050, 1181, 1300, 1386, and \SI{1440}{\per\centi\meter} are present for both sputtered films and the single-crystal substrate. They are the first overtones or combinations of the fundamental modes \cite{mccarty_1989}, which can be Raman-active even if the fundamental mode is silent.

  The combination of magnetization and Raman spectra measurements show that the $I$, $R$, and $I\!R$ symmetries are removed from the magnetically-ordered phase, and the $I$ symmetry is removed from the paramagnetic phase. The oxygen octahedra present the mirror-primed symmetry elements. Based on the relaxation of the films, we do not expect a significant distortion to the oxygen octahedra. Therefore, the MPG of our sputtered \chromia{} is only lowered from $\Bar{3}'m'$ to $3m'$. If the oxygen octahedra have vacancies, the MPG will be further lowered to $3$. In either case, quadratic and linear ME effects are allowed.

\begin{table}
    \caption{\label{tab:modes} Identified modes in Raman spectra of \chromia{} bulk and film samples. Note that the fourth mode position values cannot be determined due to the large peak of $A_{1g}$ mode, and they are for indication only.}
    \begin{ruledtabular}
    \begin{tabular}{l l l l l}
         Mode & \multicolumn{4}{l}{Raman shift [\si{\per\centi\meter}]} \\
            & Bulk & 2300-\si{\nano\meter} &1000-\si{\nano\meter} &500-\si{\nano\meter} \\
         \hline \\
        $E_g$ or $A_{1g}$ & 295.2&295.0&295.9&297.6\\
        $E_g$             &348.9&350.0&349.8&350.6\\
        $E_g$             &398.1&394.1&394.5&395.0\\ 
        $E_g$             &(528)&(520)&(520)&(521)\\
        $A_{1g}$          &552.4&553.2&550.2&549.8\\
        $E_g$             &613.7&614.5&611.0&610.2\\
    \end{tabular}
    \end{ruledtabular}
\end{table}

\section{Quadratic ME effect of \chromia{} during field cooling}
In the textured \chromia{} films, the final domain state after MEFC is governed by the short-range-ordered activation domains near $T_N$ \cite{al-mahdawi_2017a,appel_2019}. The average domain state ($\langle L \rangle$) is found from average linear ME response of the sample, normalized to the saturation value ($\langle L \rangle = \alpha_\mathrm{meas} /\alpha_\mathrm{max} $). The $\langle L \rangle$ can be defined from the Boltzmann distribution ($P_\Theta$) of the instantaneous angle ($\theta$) of $\ell$ with the $c$-axis as a function of the microscopic free energy $W$ at $T_N$, as follows \cite{al-mahdawi_2017a,al-mahdawi_2017b}:

\begin{align}
    \label{eq:L_avg_def}
    P_\Theta(\theta) \propto& \sin\theta \exp\left(\frac{-W\left(\theta\right)}{k_B T_N}\right)\,,\\
    \langle L \rangle =& \frac{\alpha_\mathrm{meas}}{\alpha_\mathrm{max}} \nonumber \\
    \equiv& \frac{\int_0^{\pi} \mathrm{sgn}\left(\cos \theta\right) P_\Theta \left( \theta \right) d\theta}{\int_0^{\pi} P_\Theta\left(\theta\right) d\theta}\,,
\end{align}
where $\mathrm{sgn}(\cos \theta)$ is the signum function which gives the opposite sign to each domain state, and $k_B$ is the Boltzmann constant.

During the MEFC, the freezing electric $E_\mathrm{fr}$ and magnetic $H_\mathrm{fr}$ fields are applied along the $c$-axis. Then a single fluctuating magnetic domain can be decomposed into $c$-axis and $ab$-plane components. The magnetic susceptibility $\chi$ is an even-rank polar tensor invariant under time reversal ($+i$ symmetry), which has a $\cos^2 \theta$ character. In the $c$-axis component, the linear ME energy and Zeeman energy terms have a $\cos \theta$ dependence. The $\beta$ tensor is an odd-rank polar tensor invariant under time reversal ($-i$ symmetry), therefore the electrobimagnetic energy term will have also a $\cos \theta$ dependence. For the linear and nonlinear ME energy terms, the $ab$-plane projections are zero due to crystal symmetry \cite{grimmer_1991,grimmer_1994}. We are using small electric fields, so we can ignore the energy terms of pyroelectricity $P E$, electric susceptibility $\epsilon E^2$, and magnetobielectricity $\gamma HE^2$. Furthermore, since $E_\mathrm{fr}$ and $H_\mathrm{fr}$ are applied along the $c$-axis, the relevant tensor terms are only $M_3$, $\chi_{33}$, $\alpha_{33}$, and $\beta_{333}$, which will refer to without indices in the following. To account for the case of an interfacial exchange-coupling with an adjacent ferromanget, such as Co, we include it using a simplistic form of an effective exchange coupling energy $J_K$ divided by \chromia{} film thickness $t$ \cite{borisov_2005,al-mahdawi_2017b}. Then, the energy $W$ for a single activation particle with a volume $V$ is as follows:

\begin{align}
    \label{eq:F-angle}
    W/V =& \frac{J_K}{t} \cos \theta -M H_\mathrm{fr} \cos \theta   \nonumber \\
    & -\alpha E_\mathrm{fr}H_\mathrm{fr} \cos \theta -\frac{1}{2} \beta E_\mathrm{fr} H_\mathrm{fr}^2 \cos \theta \nonumber\\
    & - \frac{1}{2} \chi H_\mathrm{fr}^2 \cos^2 \theta \nonumber \\
    \equiv & \, W_1 \cos \theta + W_2 \cos^2 \theta \,.
\end{align}

At the threshold condition of $\langle L \rangle = 0$, there is an equal probability of a domain to be in either a $L^+$ or a $L^-$ state. For this condition to hold, $P_\Theta$ becomes symmetric around $\theta=\pi/2$, and $dP_\Theta / d\theta \left(\theta = \pi/2\right) = 0$. It can be seen that the threshold condition is $W_1 = 0$, and the value of $W_2$ does not affect the threshold. The threshold electric field $E_\mathrm{th}$ is then found as follows:

\begin{align}
    \label{eq:Eth-JK}
    E_\mathrm{th} &=  \frac{ \frac{J_K}{t H_\mathrm{fr}} - M}{\alpha + \frac{1}{2} \beta H_\mathrm{fr}} \nonumber \\
    &\approx -\frac{1}{\alpha} \left(M - \frac{J_K}{t H_\mathrm{fr}} \right) \left(1 -  \frac{\beta}{2\alpha} H_\mathrm{fr} \right) \nonumber \\
    &\equiv A + B \cdot \frac{1}{\lvert H_\mathrm{fr} \rvert} + C \cdot \lvert H_\mathrm{fr} \rvert \, .
\end{align}

In this report, we do not use exchange-coupled films ($J_K = 0$). Therefore, Eq.~\ref{eq:Eth-JK} simplifies as follows:

\begin{align}
    \label{eq:Eth}
    E_\mathrm{th} &\approx -\frac{M}{\alpha} \left(1 -  \frac{\beta}{2\alpha} H_\mathrm{fr} \right) \nonumber \\
    &\equiv A + C \cdot \lvert H_\mathrm{fr} \rvert \, .
\end{align}
The effect of $\beta$ has a distinct qualitative feature of a linear shift in $E_\mathrm{th}$ by the applied $H_\mathrm{fr}$ during MEFC.
On the other hand, the presence of $M$ gives a constant shift in $E_\mathrm{th}$ \cite{al-mahdawi_2017b, nozaki_2018b}.  For a bulk \chromia, both $M$ and $\beta$ are zero, and $E_\mathrm{th}$ is expected to be zero for all MEFC conditions. We need to note that $\beta$ is invariant under time reversal. By inspecting the case in Fig.~\ref{fig:mpg}(b), the quadratic ME effect favors the initial parallel alignment of $M$ and $\ell$ vectors, and $\beta$ will flip its sign when $M$ changes direction by external magnetic field. Therefore, the absolute value of $H_\mathrm{fr}$ should be used in Eq.~\ref{eq:Eth}.

\begin{figure}
	\includegraphics[width=0.46\textwidth]{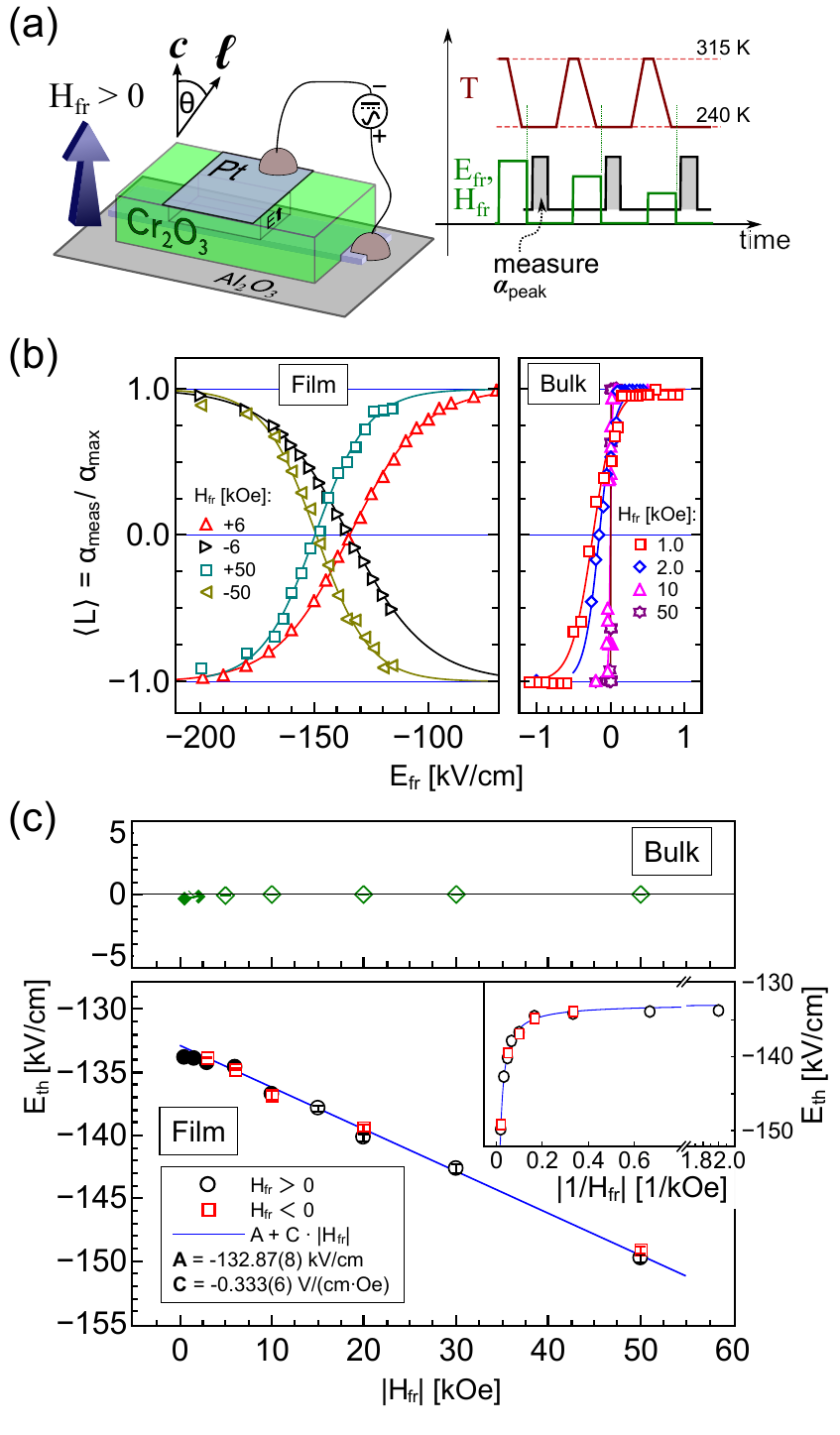}
	\caption{(a) The schematic of the film sample and the measurement procedure.
	(b) The dependence of average domain state $\langle L \rangle$ after MEFC on $E_\mathrm{fr}$ is shown with varying $\lvert H_\mathrm{fr} \rvert$. The solid lines are fits to a $\tanh$ function. (c) For the film sample, $E_\mathrm{th}$ shows a linear dependence on $H_\mathrm{fr}$, whereas $E_\mathrm{th} = 0$ for all conditions. The error bars indicate the fitting confidence, and the solid symbols are from Ref.~\cite{al-mahdawi_2017b}. The inset shows that there is no $1/H_\mathrm{fr}$ dependence. 
	}
	\label{fig:LE}
\end{figure}

Fig.~\ref{fig:LE}(a) shows the schematic of samples and measurement procedures.
To observe the quadratic ME effect, we scanned the MEFC conditions including high magnetic fields $\lvert H_\mathrm{fr} \rvert > \SI{10}{\kilo\oersted}$. The bulk sample is the reference sample with no expected high-order ME effects, and the film sample shows a small volume magnetization $M = \SI{0.2}{\emucc}$.
After setting the temperature at \SI{315}{\kelvin}, which is above $T_N$, the sample is cooled under various $E_\mathrm{fr}$ and $H_\mathrm{fr}$ down to the $\alpha$ peak temperature. Then $\langle L \rangle$ is found by normalizing $\alpha_\mathrm{peak}$ to its maximum value. For a fixed $H_\mathrm{fr}$, $E_\mathrm{th}$ is determined from fitting $\langle L \rangle$ to a hyperpolic tangent function \cite{al-mahdawi_2017b}.
The positive directions of $E_\mathrm{fr}$ and $H_\mathrm{fr}$ are along the magnetometer's positive sensing direction, as indicated in Fig.~\ref{fig:LE}(a).

Fig.~\ref{fig:LE} shows the experimental results. After sweeping $E_\mathrm{fr}$ to be parallel or anti-parallel to $H_\mathrm{fr}$ direction, the domain state shows a transition from $+1$ ($L^+$ state) to $-1$ ($L^-$ state) in both samples [Fig.~\ref{fig:LE}(a)]. The crossing of $\langle L \rangle = 0$ occurs at the threshold condition of $E_\mathrm{fr} = E_\mathrm{th}$, as defined in Eq.~\ref{eq:Eth}. In the bulk sample, $E_\mathrm{th} \approx 0$ for all applied $H_\mathrm{fr}$ values, as expected due to the lack of $M$ and $\beta$. On the other hand, in the film sample, $E_\mathrm{th}$ has a constant shift $A < 0$, due to the parallel alignment between $M$ and $\ell$ as shown in Fig.~\ref{fig:mpg}(b) \cite{al-mahdawi_2017b,nozaki_2017a}. Furthermore, $E_\mathrm{th}$ shows a shift towards the negative side as $\lvert H_\mathrm{fr} \rvert$ is increased. The dependence of $E_\mathrm{th}$ on $\lvert H_\mathrm{fr} \rvert$ is plotted in Fig.~\ref{fig:LE}(b), with the complementary $E_\mathrm{th} - \lvert 1/H_\mathrm{fr} \rvert$ dependence shown in the inset. We only find a linear dependence of $E_\mathrm{th}$-$\lvert H_\mathrm{fr} \rvert$ line, that is symmetric for $\pm H_\mathrm{fr}$, as expected from a quadratic ME effect, as shown in Eq.~\ref{eq:Eth}. The quadratic ME energy prefers the initial alignment of $M\parallel \ell$, and it opposes the linear ME energy during $-$MEFC. Therefore a larger negative electric field is required to switch the domain state at higher magnetic fields.

An estimation of quadratic ME susceptibility relative to the linear ME effect can be found from the slope ($C$) and intercept ($A$) of $E_\mathrm{th}$-$\lvert H_\mathrm{fr} \rvert$ line. Using the fitting values in Fig.~\ref{fig:LE}(b), $\beta / \alpha \approx -2 \cdot C/A = \SI{-5.0e-6}{\per\oersted}$. The value of $\alpha(T_N)$ in Eq.~\ref{eq:F-angle} is the microscopic value, which is larger than the macroscopically-observed peak value of $\alpha$ \cite{al-mahdawi_2017b,ahmed_2018,shiratsuchi_2020a}. We estimate $\beta =   \SI{-1.69e-18}{\second\per\ampere}$, based on our previous estimation of microscopic $\alpha = \SI{27e-12}{\second\per\meter}$ \cite{al-mahdawi_2017b}.
In other multiferroic materials, such as NiSO$_4 \cdot 6$H$_2$O \cite{hou_1965} and BiFeO$_3$ \cite{tabares-munoz_1985}, the magnitude of $\beta$ is comparable to our observation, albeit reported at low temperatures of 3--4 K.
Other multiferroics with a non-collinear spin structure have a large quadratic ME effect, observed at very low temperatures and high fields \cite{fogh_2020,weymann_2020,kimura_2012,chai_2014,popov_2020}. The microscopic mechanism of the quadratic ME effect in a collinear ferrimagnet requires further investigation. In the present \chromia{} system, it is likely related to the same mechanism of the linear ME effect, namely the two-spin symmetric exchange interaction \cite{mostovoy_2010,matsumoto_2017}, in combination with the removal of $I\!R$ symmetry. 

\section{Conclusions}
 We investigated a quadratic ME effect, also called the electrobimagnetic and paramagnetoelectric effects, in sputtered \chromia{} films. We found that the quadratic ME energy contributed to the ME switching during the field cooling process. The quadratic ME effects are forbidden in the magnetic point group of bulk \chromia{}.
 The emergence of this effect in films is due to removal of space-time inversion symmetry. The magnetization and Raman spectra measurements show that the space, time, and space-time symmetries are removed, likely due to site-selective doping of trace dopants in \chromia{} films during growth.

Further applications to the quadratic ME effects might be found in optical rectification and frequency multiplication \cite{schmid_1973,kamentsev_2006,saito_2009}. The magnetic point group of $3m'$ is one of the thirteen groups that allow both spontaneous electric and magnetic polarizations in addition to various ME effects \cite{schmid_2008,borovik-romanov_2013,ye_2016}. Therefore, we suggest that investigating and engineering the linear and quadratic ME effect in \chromia{} films open new ways of ME control in spintronic devices.

\begin{acknowledgments}
The authors thank M. Nemoto (Technical Division of Tohoku University) for the technical support of Raman spectroscopic measurement.
This work was partially supported by the Center for Science and Innovation in Spintronics (CSIS), and
Center for Spintronics Research Network (CSRN), Tohoku University, and the ImPACT Program of the Council for Science, Technology and Innovation (Cabinet Office, Government of Japan).
\end{acknowledgments}

\newcommand{\noopsort}[1]{}
%

\newpage
\clearpage
\onecolumngrid
\begin{center}
\textbf{\large Supplemental Material: Quadratic magnetoelectric effect during field cooling in sputter grown Cr$_2$O$_3$ films}
\end{center}
\setcounter{section}{0}
\setcounter{equation}{0}
\setcounter{figure}{0}
\setcounter{table}{0}
\setcounter{page}{1}
\makeatletter
\renewcommand{\thesection}{S\arabic{section}}
\renewcommand{\theequation}{S\arabic{equation}}
\renewcommand{\thefigure}{S\arabic{figure}}
\renewcommand{\bibnumfmt}[1]{[S#1]}
\renewcommand{\citenumfont}[1]{S#1}
\section{Simulation of xray diffraction patterns}

\begin{figure}[ht!]
 \vspace{1\baselineskip}
	\includegraphics[width=0.85\textwidth]{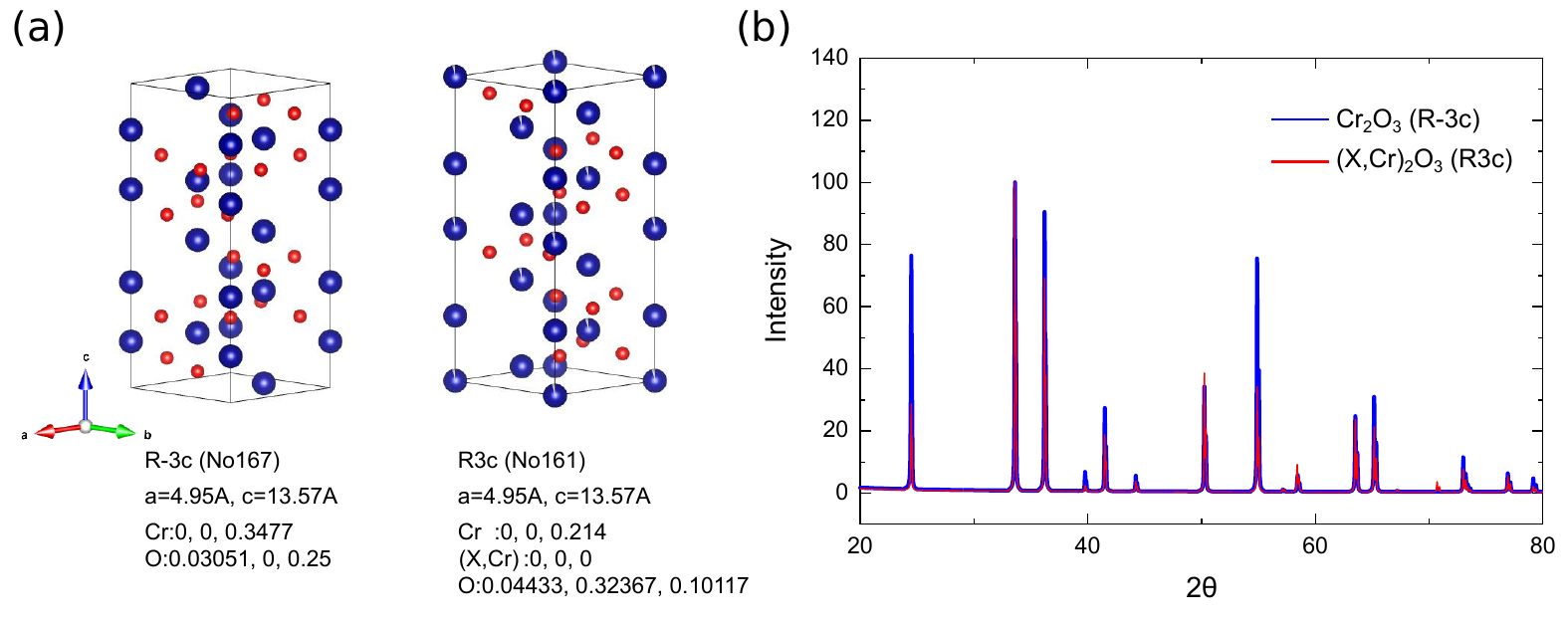}
	\caption{(a) The simulated structures of bulk \chromia{}, and doped \chromia{}.  (b) Simulated XRD profiles show that the patterns are difficult to distinguish.}
	\label{fig:xrd}
\end{figure}

We show the effect of site selective doping in \chromia{} on diffraction patterns using simulation. We simulate the xray powder diffraction patterns (XRD), using \textsc{vesta} and \textsc{rietan-fp} software packages \cite{vesta_supp,RIETAN-FP_supp}. We use the two depicted crystal structures in Fig.~\ref{fig:xrd}(a). We keep the same lattice constants of bulk \chromia{} for each, but we change the atom species on a single Cr sublattice to (X,Cr), to represent the site-selective substitution. Thus, we simulate the crystallographic space groups Nos.~167 ($R\Bar{3}c$), and 161 ($R3c$).

We find that the XRD peaks positions are the same between the two structures [Fig.~\ref{fig:xrd}(b)]. The Friedel's law dictates that noncentrosymmetric and centrosymmetric crystals have the same symmetric xray (and neutron) diffraction spectra. Anomalous xray scattering may change the intensity ratios in a noncentrysmmetric crystal, as seen in Fig.~\ref{fig:xrd}(b). However, the differences are within an order of magnitude. Experimentally in thin films, there will be other distortion effects on the XRD peaks, and it is difficult to distinguish the removal of space-inversion symmetry using diffraction methods.

\end{document}